\documentclass[journal]{IEEEtran}
\ifCLASSINFOpdf
\else
   \usepackage{graphicx}
   \graphicspath{{../eps/}}
   \DeclareGraphicsExtensions{.eps}
\fi
\usepackage{fancyhdr}
\usepackage{amsmath,amssymb,amstext}
\usepackage{subeqnarray}
\usepackage{cite}
\usepackage{hhline}
\usepackage{bm}
\usepackage{soul}
\usepackage{slashbox}
\usepackage{times,amsmath,amsfonts}
\usepackage{subfigure}
\usepackage{amsmath,amssymb,bm}
\usepackage{graphicx,picture}
\usepackage{cite}
\usepackage{subfigure}
\usepackage{subeqnarray}
\usepackage{cases}
\usepackage{color}
\usepackage{overpic}
\usepackage{multirow}
\usepackage{booktabs}
\usepackage{epstopdf}
\usepackage[noend]{algpseudocode}
\usepackage{algorithmicx,algorithm}

\definecolor{r}{rgb}{1,0,0}
\definecolor{b}{rgb}{0,0,1}
\definecolor{k}{rgb}{0,1,1}

\usepackage{upgreek}

\newcounter{saveeqn}%

\allowdisplaybreaks
\DeclareMathSymbol{\Phi}{\mathord}{letters}{8}

\begin{document}
\title{Rethinking Waveform for 6G: Harnessing Delay-Doppler Alignment Modulation}

\author{
\IEEEauthorblockN{Zhiqiang Xiao, Xianda Liu, Yong Zeng~\IEEEmembership{Senior Member, IEEE}, \\
J. Andrew Zhang~\IEEEmembership{Senior Member, IEEE}, Shi Jin~\IEEEmembership{Fellow, IEEE}, and Rui Zhang~\IEEEmembership{Fellow, IEEE}}

\thanks{
Z. Xiao, X. Liu, Y. Zeng, and S. Jin are with Southeast University, China; Z. Xiao and Y. Zeng are also with the Purple Mountain Laboratories, China. (e-mail: \{zhiqiang\_xiao, 220230853, yong\_zeng, jinshi\}@seu.edu.cn). (\emph{Corresponding author: Yong Zeng.})

J. Andrew Zhang is with University of Technology Sydney, Australia. (e-mail: Andrew.Zhang@uts.edu.au)

Rui Zhang is with The Chinese University of Hong Kong, Shenzhen, Shenzhen Research Institute of Big Data, and National University of Singapore. (e-mail: elezhang@nus.edu.sg)
}
}

\maketitle
\begin{abstract}
Waveform design has served as a cornerstone for each generation of mobile communication systems.
The future sixth-generation (6G) mobile communication networks are expected to employ larger-scale antenna arrays and exploit higher-frequency bands for further boosting data transmission rate and providing ubiquitous wireless sensing.
This brings new opportunities and challenges for 6G waveform design.
In this article, by leveraging the super spatial resolution of large antenna arrays and the multi-path spatial sparsity of high-frequency wireless channels, we introduce a new approach for waveform design
based on the recently proposed delay-Doppler alignment modulation (DDAM).
In particular, DDAM makes a paradigm shift of waveform design from the conventional manner of tolerating channel delay and Doppler spreads to actively manipulating them.
First, we review the fundamental constraints and performance limitations of orthogonal frequency division multiplexing (OFDM) and introduce new opportunities for 6G waveform design.
Next, the motivations and basic principles of DDAM are presented, followed by its various extensions to different wireless system setups.
Finally, the main design considerations for DDAM are discussed and the new opportunities for future research are highlighted.
\end{abstract}

\section{Introduction}
The essence of waveform design is to synthesize transmit waveforms for conveying the information-bearing symbols most efficiently over the propagation channels.
For wireless communication, transmitted signals usually undergo multi-path propagation and arrive at the receiver with different delays.
Typically, if the symbol duration is smaller or even comparable to the channel delay spread, the detrimental inter-symbol interference (ISI) occurs.

Over the past few decades, all generations of mobile communication systems have witnessed a continuous effort
on waveform design to mitigate the ISI.
In the second-generation (2G) and third-generation (3G) eras, due to the relative small signal bandwidth, single-carrier waveforms combined with time-domain equalization (TEQ) or spread spectrum and RAKE receiver were the main techniques for ISI mitigation.
For the fourth-generation (4G) and fifth-generation (5G) mobile communication networks, orthogonal frequency division multiplexing (OFDM) made a great success, which can convert the frequency-selective channel to parallel frequency-flat sub-channels via multi-carrier transmission.
Thus, it can effectively mitigate the ISI and allow for flexible time-frequency (TF) resource allocation \cite{cho2010mimo}.
However, OFDM also suffers from practical issues, such as large cyclic prefix (CP) overhead, high peak-to-average power ratio (PAPR), severe inter-carrier inference (ICI) in high-mobility scenarios, and high out-of-band emission (OOBE) \cite{cho2010mimo}.
Although numerous designs have been proposed to cope with such issues, like discrete Fourier transform
spread OFDM (DFT-s-OFDM) to reduce the PAPR, and filter band multi-carrier (FBMC) or generalized frequency
division multiplexing (GFDM) to suppress the OOBE and improve the spectral efficiency (SE) \cite{sahin2013survey}, they either increase the implementation complexity or degrade the system
performance.
Thus, waveform design for broadband mobile communication is still an important problem not fully solved.

The sixth-generation (6G) mobile communication networks are expected to not only further enhance the communication
performance, but also support a lot of new scenarios and services, such as integrated sensing and communication (ISAC) \cite{ITU-R} and space-air-ground integrated (SAGI) communications \cite{you2021towards}.
Thus, waveform design needs to consider wireless sensing and high-mobility support capabilities.
To this end, delay-Doppler (DD) domain waveform design has drawn increasing attentions recently, due to its robustness against channel fluctuations in high-mobility scenarios and its potential for high-performance ISAC.
Two typical examples for DD domain waveforms are orthogonal time frequency space (OTFS) \cite{hadani2017orthogonal} and orthogonal delay-Doppler modulation (ODDM) \cite{lin2022orthogonal}.
In general, wireless channels suffer from time-frequency doubly selective (TFDS) fading.
The main idea of OTFS is to convert each DD tap into the entire TF plane for exploiting all multi-path diversities, which can be similarly achieved via vector OFDM (VOFDM) \cite{xia2001vOFDM}.
This makes OTFS (or VOFDM) superior than OFDM over TFDS fading channels.
However, as demonstrated in \cite{xia2024delay}, for single-antenna systems, neither OTFS nor VOFDM can effectively compensate the channel delay and Doppler spreads, as they are coupled with all the multi-paths.

\begin{figure*}
    \centering
     \includegraphics[width=0.85\textwidth]{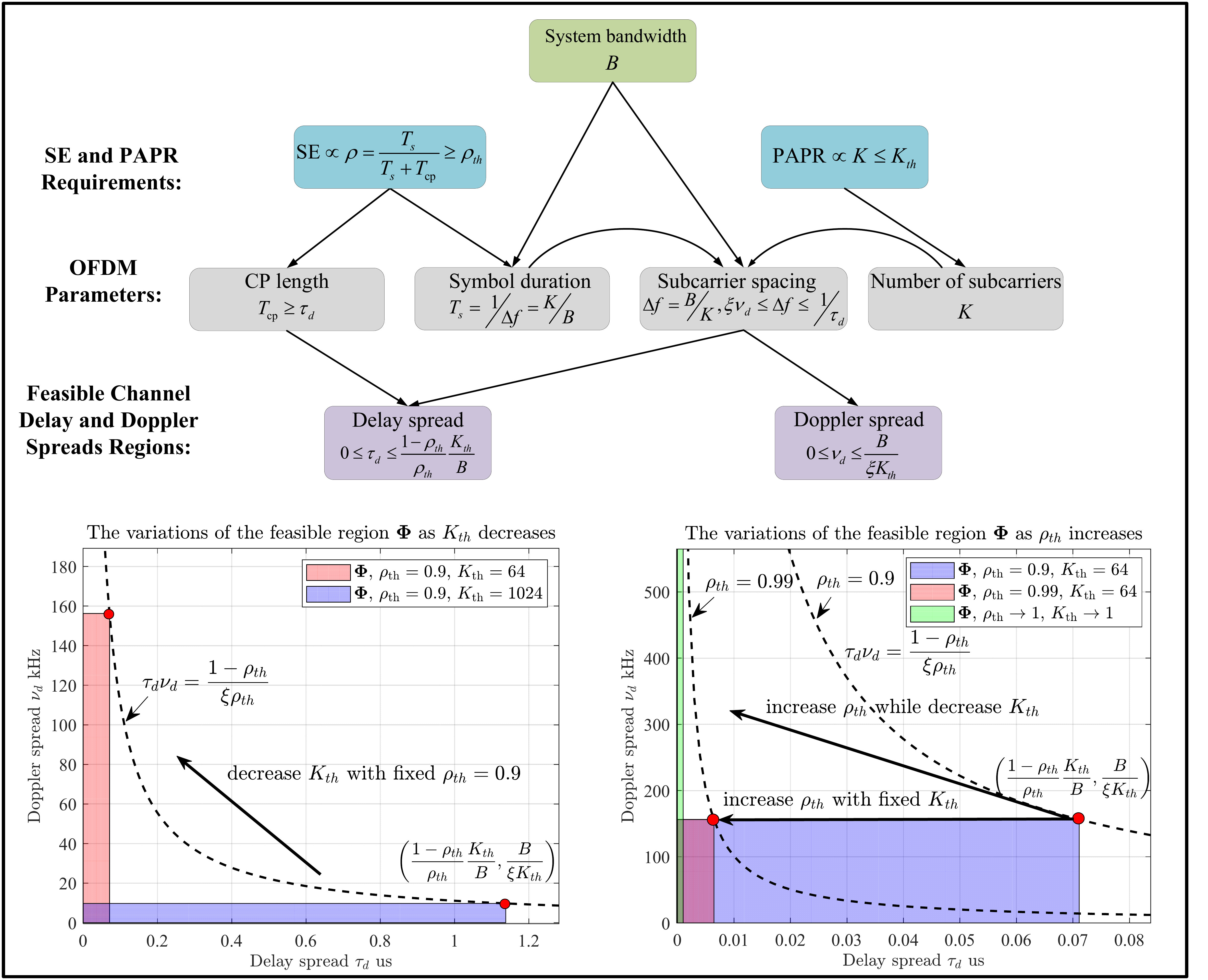}
  \caption{The relationship between the OFDM system performance requirements, its parameters, and channel delay and Doppler spreads, and the feasible region $\bm\Phi$ of channel delay and Doppler spreads to achieve the desired performance of SE and PAPR under different $(\rho_{\mathrm{th}},K_{\mathrm{th}})$ pairs.}\label{OFDMPara}\vspace{-10pt}
\end{figure*}
High-frequency transmission at millimeter wave (mmWave) or Terahertz (THz) band and extremely large-scale multiple-input multiple-output (XL-MIMO) stand out as two promising techniques for 6G, which renders wireless channels much sparser than conventional sub-6G channels.
Besides, mmWave XL-MIMO systems have high resolution and abundant design degrees-of-freedom (DoFs) in the spatial domain.
As such, not only the composite channels of different users, but also the individual channel paths of the same user, could be resolved in the spatial domain.
This thus motivated the recently proposed delay-Doppler alignment modulation (DDAM) technique \cite{lu2023delay}, whose key ideas are to resolve all the channel multi-paths in the spatial domain via per-path based beamforming and coherently combine the signals over these resolved multi-paths after properly compensating their delays and Dopplers.
By doing so, the resulting channel delay and Doppler spreads can be significantly reduced or even completely eliminated.
Thus, DDAM brings an unprecedented opportunity for a paradigm shift in future waveform design from the conventional approach of passively {\it tolerating} the channel delay and Doppler spreads to {\it manipulating} them in a proactive way.

The objective of this article is to introduce a new framework for waveform design, namely {\it DDAM + ``X''}, whereby DDAM can be combined with various waveform design techniques such as OFDM and OTFS to improve their performance.
To this end, the fundamental constraints and performance limitations of OFDM systems are first reviewed in Section~\ref{OFDM constraints}, followed by the new opportunities for 6G waveform design in Section~\ref{trend}.
In Section~\ref{DDAM_main}, we present the principles of DDAM and various ways to combine DDAM with other waveforms, including DDAM-OFDM and DDAM-OTFS, with their main design considerations addressed.
Several future research directions are then discussed in Section~\ref{directions}.
Finally, we conclude this article in Section~\ref{clu}.

\section{Fundamental Constraints and Performance Limitations of OFDM}\label{OFDM constraints}

To motivate our subsequent discussions, we first consider the fundamental constraints on the parameter selection for OFDM systems\footnote{Note that the similar analysis can be also derived for OTFS.}.
The system bandwidth is $B$ and the subcarrier spacing is $\triangle f = \frac{B}{K}$, with $K$ subcarriers.
The OFDM symbol duration without CP is $T_s=\frac{1}{\triangle f}=\frac{K}{B}$.
Consider a time-varying channel with delay spread $\tau_d$ and Doppler spread $\nu_d$.
To ensure the orthogonality of the received signal, the CP length, denoted by $T_{\mathrm{cp}}$, should be no smaller than the delay spread, i.e., $T_{\mathrm{cp}}\ge\tau_d$, while $\triangle f$ should be much greater than the Doppler spread, i.e., $\triangle f\ge \xi \nu_d$, where $\xi$ is a scaling factor with a typical value $\xi=10$ \cite{cho2010mimo}.
Moreover, to ensure the frequency flatness within each subcarrier, $\triangle f $ should be no greater than the channel coherence bandwidth $B_c$, i.e., $\triangle f\le B_c$, where $B_c\approx 1/\tau_d$ \cite{cho2010mimo}.
Thus, the subcarrier spacing $\triangle f$ should satisfy that $\xi\nu_d\le \triangle f\le 1/\tau_d$.
Under the above setup, the following question is raised: What are the feasible regions of channel delay and Doppler spreads $(\tau_d,\nu_d)$ under which a given performance pair of  SE and PAPR can be achieved by OFDM?

\begin{figure*}
    \centering
     \includegraphics[width=0.9\textwidth]{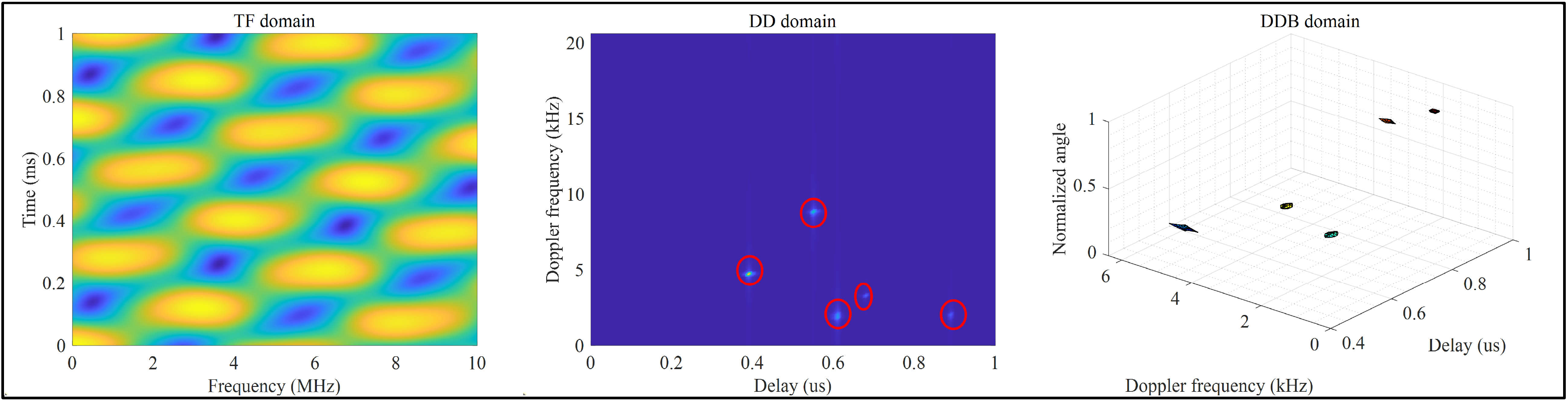}
  \caption{The time-varying channel consisting of $L=5$ multi-paths expressed in TF domain, DD domain, and DDB domain.}\label{ChannelDomain}\vspace{-10pt}
\end{figure*}
Note that while SE and PAPR depend on various factors, to gain the essential insights, we use the CP ratio $\rho=\frac{T_s}{T_s+T_{\mathrm{cp}}}$ and the number of OFDM subcarriers $K$ as their performance indicators, respectively.
From Fig.~\ref{OFDMPara}, it follows that to achieve $\rho\ge\rho_{\mathrm{th}}$ and $K\le K_{\mathrm{th}}$, with $\rho_{\mathrm{th}}$ and $K_{\mathrm{th}}$ being the given thresholds, the delay and Doppler spreads should lie within the region $\bm\Phi\triangleq\{(\tau_d,\nu_d)\mid 0\le \tau_d\le\frac{1-\rho_{\mathrm{th}}}{\rho_{\mathrm{th}}}\frac{K_{\mathrm{th}}}{B}, 0\le \nu_d\le\frac{B}{\xi K_{\mathrm{th}}}\}$,
which is a rectangle for a given $(\rho_{\mathrm{th}},K_{\mathrm{th}})$ pair.
However, in practice, the delay and Doppler spreads of the natural wireless channels may not lie in the feasible region $\bm\Phi$.
Specifically, as shown in Fig.~\ref{OFDMPara}, for a given $\rho_{\mathrm{th}}$, the feasible delay spread decreases with $K_{\mathrm{th}}$.
On the other hand, to increase $\rho_{\mathrm{th}}$ while maintaining $K_{\mathrm{th}}$, it also requires to reduce the delay spread.
Moreover, the size of the feasible region $\bm\Phi$ decreases as $\rho_{\mathrm{th}}$ increases.
This is because for any given $\rho_{\mathrm{th}}$, we have $\tau_d\nu_d\le \frac{1-\rho_{\mathrm{th}}}{\xi\rho_{\mathrm{th}}}$, as illustrated in the black dashed lines in Fig.~\ref{OFDMPara}.
In particular, to satisfy the extreme SE and PAPR performance, i.e., $\rho_{\mathrm{th}}\rightarrow 1$ and $K_{\mathrm{th}}\rightarrow 1$, as shown in Fig.~\ref{OFDMPara}, the delay spread needs to be reduced to nearly zero.

Some prior works were devoted to mitigating the channel delay and Doppler spreads.
Specifically, to mitigate the channel delay spread and the resulted ISI, many TEQ methods were proposed, such as channel shortening \cite{melsa1996impulse} and time-reversal (TR) \cite{wang2011green}.
However, the channel delay spread and ISI cannot be completely eliminated.
On the other hand, while carrier frequency offset (CFO) compensation is commonly employed to mitigate the Doppler frequency shift in OFDM systems, it is difficult to effectively mitigate the composite Doppler frequency shift of all multi-paths \cite{cho2010mimo}.
Looking ahead to 6G waveform design, it still remains open if the channel delay and/or Doppler spreads can be manipulated flexibly to achieve the extreme performance of $\rho_{\mathrm{th}}\rightarrow 1$ and $K_{\mathrm{th}}\rightarrow 1$.

\section{New Opportunities for 6G Waveform Design}\label{trend}
In this section, we introduce some new opportunities for 6G waveform design.

{\it \textbf{Multi-path spatial sparsity}}:
6G is likely to further exploit higher frequency spectrum, such as mmWave and THz.
With the increase of transmission frequencies, the number of multi-paths will be significantly reduced.
Thus, different from conventional rich scattering sub-6G channels, multi-path channels in mmWave/THz bands will exhibit spatial sparsity.
Note that this does not imply that the channel delay or Doppler spread will become small, as the differences in delay or Doppler among the less multi-paths may be still large.
By leveraging the multi-path spatial sparsity, it is more feasible to distinguish multi-path components for path-based signal processing to mitigate the channel delay and Doppler spreads.

{\it \textbf{High spatial resolution}}:
The scale of antennas is expected to increase further, from tens of elements in 5G to hundreds or even thousands of them in 6G.
This will provide unprecedentedly high spatial resolution and abundant DoFs in the spatial domain, benefiting both wireless communication and sensing.
By leveraging the super-high spatial resolution enabled by large-scale antenna arrays, the asymptotic channel orthogonality can be achieved not only among multi-users but also among multi-path channels for the same user \cite{lu2023delay}.
This brings new opportunities to resolve multi-paths in the spatial domain.

{\it \textbf{ISAC}}:
ISAC has been identified as one of the six major usage scenarios of 6G \cite{ITU-R}.
In conjunction with mmWave/THz transmission and XL-MIMO, communication and sensing will share common wireless channels in the future.
Specifically, as illustrated in Fig.~\ref{ChannelDomain}, wireless channels for communications are usually modelled in the TF domain, which may suffer from TFDS fading.
In contrast, with the multi-path spatial sparsity, wireless channels can be also represented by only a few channel coefficients in the DD domain.
By further leveraging the high spatial resolution in 6G, channel characteristics can be captured by the states of the major scatterers in the propagation environment, such as delays, Doppler frequencies, and normalized angles, which can be represented in the delay-Doppler-beamspace (DDB) domain.
Consequently, it becomes possible for unifying environment sensing and channel estimation, to extract the features of individual multi-path for both communication and sensing, rather than estimating the composite channel coefficients from all the multi-paths.

\begin{figure*}
  \centering

  \includegraphics[width=0.95\textwidth]{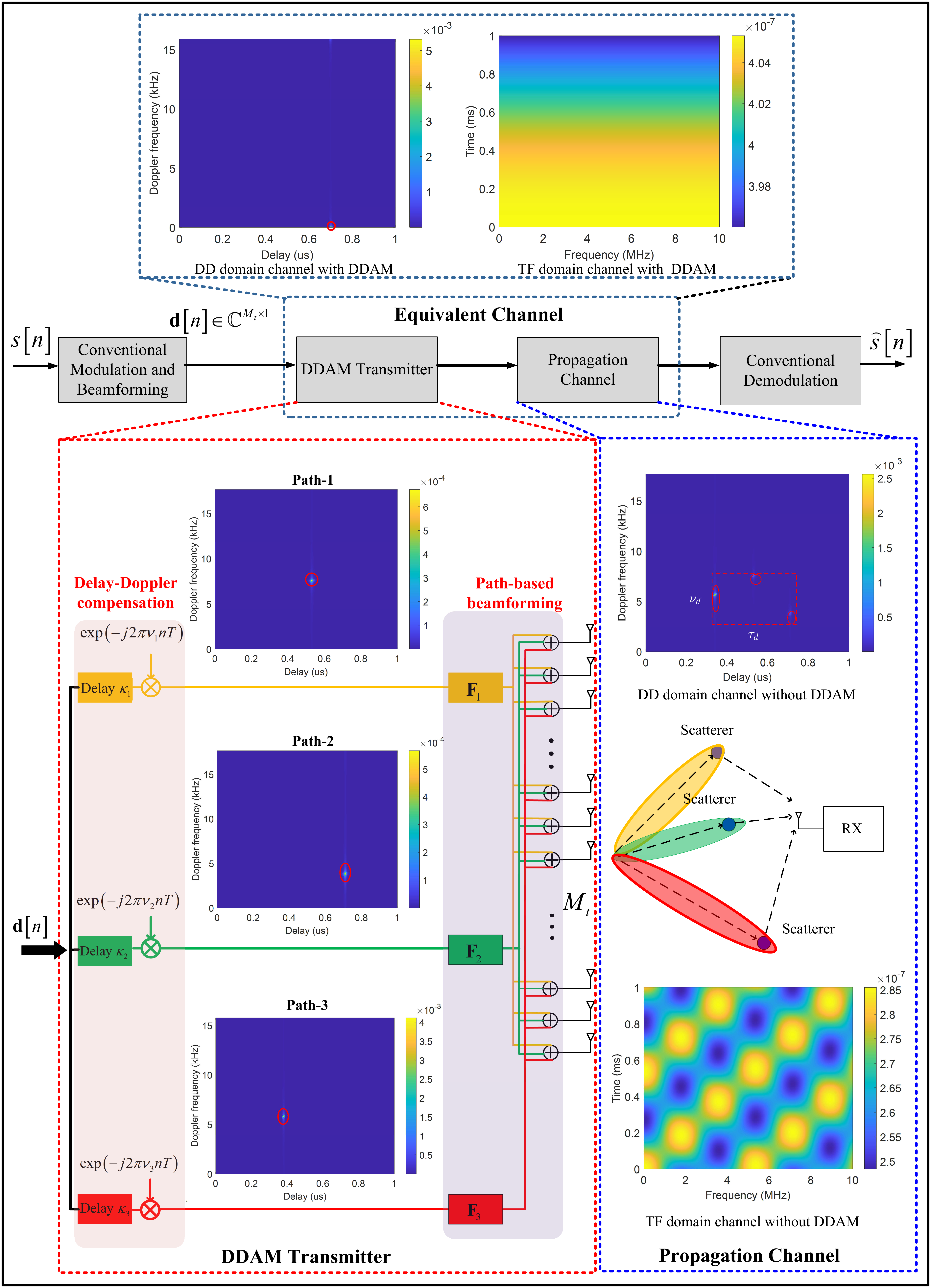}
  \caption{The framework of DDAM transmitter and its principles to manipulate the channel delay and Doppler spreads. The composite multi-path channel can be decomposed into sub-channels for each path through path-based beamforming and superimposed after delay-Doppler compensation, resulting an equivalent quasi-static flat fading channel. }\label{DDAM}\vspace{-10pt}
\end{figure*}
\section{Manipulating Channel Spread via DDAM}\label{DDAM_main}
By leveraging the multi-path sparsity and the abundant design DoFs in the spatial domain, a promising technique termed DDAM was recently proposed \cite{lu2023delay}, which can significantly reduce the channel delay and Doppler spreads, while making full use of the multi-path diversity.
It brings a new paradigm shift in waveform design, from passively {\it tolerating} to actively {\it manipulating} the channel delay and Doppler spreads.

\subsection{Principles of DDAM}\label{pDDAM}
The DDAM technique is a generalization of the {\it delay alignment modulation} (DAM) \cite{lu2023delay2}.
Compared to DAM for time-invariant frequency-selective channels only, DDAM aims to transform the general TFDS fading channel into an equivalent quasi-static flat-fading channel through {\it delay-Doppler compensation} and {\it path-based beamforming} \cite{lu2023delay}.
Its key ideas can be traced back to our previous work on lens MIMO systems \cite{zeng2016lens}.
The DDAM technique can be applied at both transmitter and receiver sides \cite{lu2023delay}.
Here, for ease of illustration, we only consider utilizing it at the transmitter side for multiple-input-single-output (MISO) systems.

As an illustrative example, by considering the multi-path spatial sparsity of mmWave/THz channels, we assume the propagation channel comprising three multi-paths, as shown in Fig.~\ref{DDAM}, where each path has different delays and Dopplers.
As shown in Fig.~\ref{DDAM}, denote by $s[n]$ and $\hat{s}[n]$ the transmit and detected information-bearing symbols, and $\mathbf{d}[n]\in\mathbb{C}^{M_t\times1}$ the transmit signals after beamforming.
With DDAM, by leveraging the high spatial resolution brought by large-scale antenna arrays, each path can be resolved in the spatial domain with a distinct angle-of-departure (AoD).
As illustrated in Fig.~\ref{DDAM}, by exploiting the abundant design DoFs in the spatial domain, the propagation channel superimposed by all multi-paths can be decomposed into multiple sub-channels for each path via path-based beamforming.
This can be achieved by designing the beamforming matrix $\mathbf{F}_l\in\mathbb{C}^{M_t\times M_t}$ for each path $l$, according to different criteria, such as zero-forcing (ZF), regularized ZF (RZF), maximal ratio transmission (MRT), and minimum mean squared error (MMSE) \cite{lu2023delay,lu2023delay2}.

For the decomposed sub-channel for each path $l$, the delay compensation $\kappa_l$ is designed to align the propagation delay of the $l$th path to the maximum delay among all the multi-paths, while the Doppler compensation $\nu_l$ is designed to eliminate its Doppler frequency shift.
Note that the essence of delay-Doppler compensation is a time shift and phase rotation of the transmit symbol sequence.
By doing so, the processed sub-channels for all multi-paths can be more efficiently superimposed, resulting in an equivalent channel that utilizes all multi-path diversity with the reduced channel delay and Doppler spreads as evident from Fig.~\ref{DDAM}, which enjoys a quasi-static flat-fading in the TF domain.
Note that for the ideal case, if all the multi-path delays are perfectly aligned to the maximum delay while the Doppler frequency of each multi-path is completely eliminated,  an ISI-free transmission can be achieved even with a simple single-carrier transmission, by designing the path-based beamforming to null the undesired multi-path components.

Compared to OFDM and OTFS, DDAM enjoys the following advantages.

\begin{figure}
  \centering
  \includegraphics[width=0.45\textwidth]{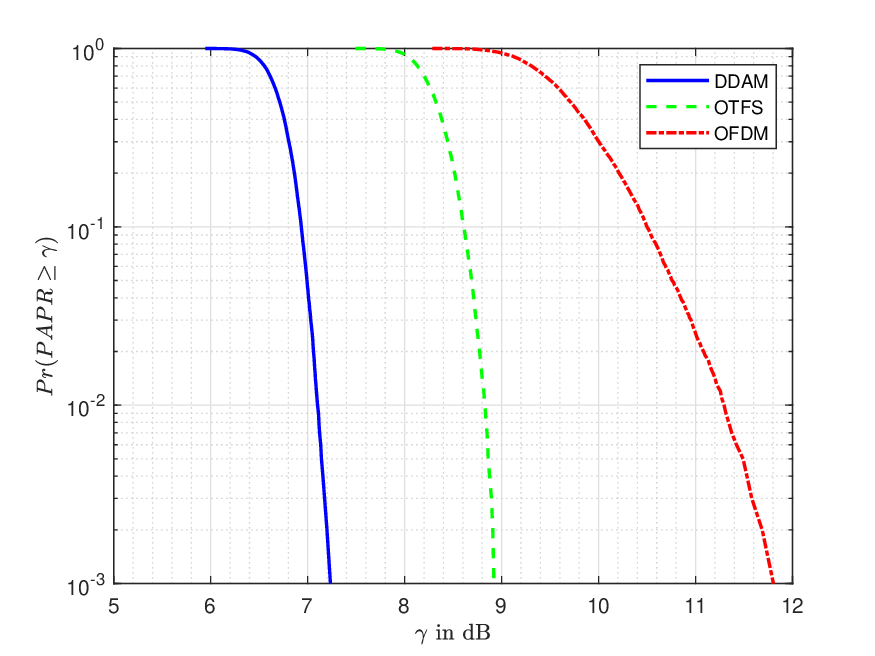}
  \caption{Comparisons of DDAM, OTFS, and OFDM on PAPR.
  }\label{PAPR}\vspace{-10pt}
\end{figure}

{\it \textbf{Low PAPR}}: The PAPR of DDAM is proportional to the number of multi-paths $L$ \cite{xiao2022waveform}, while that for OFDM or OTFS is proportional to the number of subcarriers $K$ or the number of symbols $M$, respectively.
As high-frequency wireless channels experience multi-path sparsity, it is expected that $L\ll K$ and $L<M$.
Thus, DDAM enjoys lower PAPR than OFDM and OTFS, as illustrated in Fig.~\ref{PAPR}.

\begin{figure}
  \centering
  \includegraphics[width=0.45\textwidth]{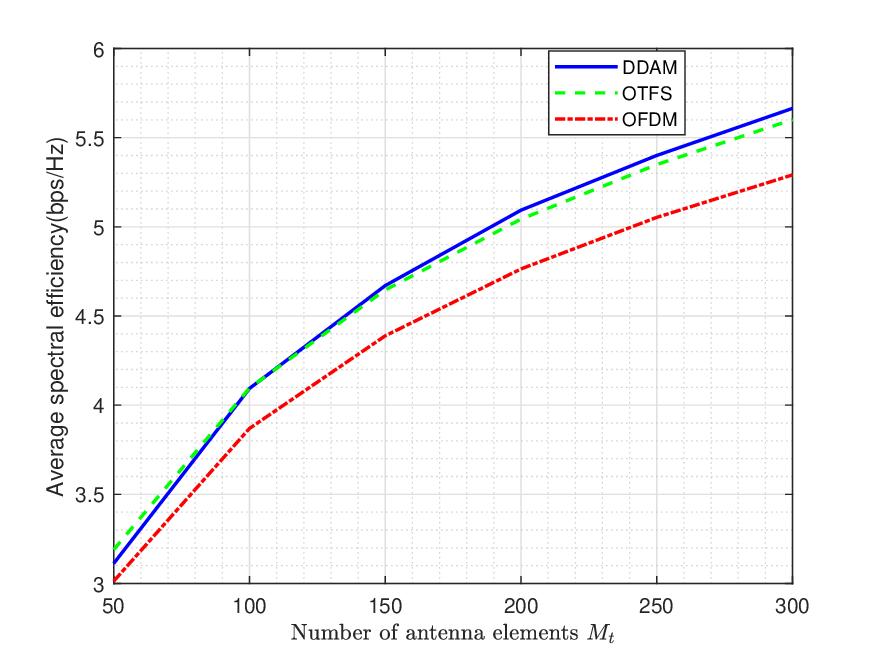}
  \caption{Comparisons of DDAM, OTFS, and OFDM on SE.}\label{SE}\vspace{-10pt}
\end{figure}

{\it \textbf{High SE}}: For DDAM, only a guard interval with the length of twice of the maximum channel delay is necessary per block to avoid the inter-block interference (IBI) \cite{lu2023delay}.
By contrast, OFDM requires the CP per symbol while OTFS requires the CP per frame to avoid the ISI.
Thus, DDAM has higher SE compared to OFDM and OTFS, as illustrated in Fig.~\ref{SE}.

{\it \textbf{Low complexity and small latency}}: DDAM can be implemented with low complexity.
The delay-Doppler compensation only involves a time shift and phase rotation of the transmit symbol sequence, while the path-based beamforming is similar to the conventional beamforming.
Different from OTFS that requires high complexity and signal processing latency, DDAM performs the simple {\it symbol-wise} signal detection at the receiver side.
Thus, DDAM enjoys low complexity and small communication latency.
In Table I, we compare the computational complexity of OFDM, OTFS, and DDAM at both transmitter and receiver sides, where the complexity is defined as the required number of multiplication for each information-bearing symbol.
It is observed that DDAM has lower complexity at both transmitter and receiver side, when the number of multi-paths is small and the number of total transmit information-bearing symbols per channel coherence block is large.
\begin{table*}[]
\footnotesize
\caption{Comparison on the complexity of OFDM, OTFS, and DDAM.}\label{Complexity comparison}
\resizebox{\textwidth}{!}{
\begin{tabular}{|cc|lcc|lcc|}
\hline
\multicolumn{2}{|c|}{\multirow{2}{*}{Waveform}}                            & \multicolumn{3}{c|}{TX}                                                                                                                                                                          & \multicolumn{3}{c|}{RX}                                                                                                                                                              \\ \cline{3-8}
\multicolumn{2}{|c|}{}                                                     & \multicolumn{1}{l|}{Operation}  & \multicolumn{1}{c|}{Complexity}                                           & Total                                                                                   & \multicolumn{1}{l|}{Operation}      & \multicolumn{1}{c|}{Complexity}                               & Total                                                                               \\ \hline
\multicolumn{2}{|c|}{\multirow{2}{*}{OFDM}}                                & \multicolumn{1}{l|}{IFFT}  & \multicolumn{1}{c|}{$\mathcal{O}(M_t\log_2K)$}     & \multirow{2}{*}{$\mathcal{O}(M_t\log_2K)$}                       & \multicolumn{1}{l|}{FFT}       & \multicolumn{1}{c|}{$\mathcal{O}(\log_2K)$}     & \multirow{2}{*}{$\mathcal{O}(\log_2K)$}                           \\ \cline{3-4} \cline{6-7}
\multicolumn{2}{|c|}{}                                                     & \multicolumn{1}{l|}{MRT}   & \multicolumn{1}{c|}{$\mathcal{O}(M_t)$}                      &                                                                                         & \multicolumn{1}{l|}{Detection} & \multicolumn{1}{c|}{$\mathcal{O}(1)$}                       &                                                                                     \\ \hline
\multicolumn{1}{|c|}{\multirow{5}{*}{OTFS}} & \multirow{3}{*}{ISFFT-based} & \multicolumn{1}{l|}{ISFFT} & \multicolumn{1}{c|}{$\mathcal{O}(M_t\log_2KM)$} & \multirow{3}{*}{$\mathcal{O}(M_t\log_2KM)$} & \multicolumn{1}{l|}{FFT}       & \multicolumn{1}{c|}{$\mathcal{O}(\log_2K)$}  & \multirow{3}{*}{$\mathcal{O}(\log_2KM)$} \\ \cline{3-4} \cline{6-7}
\multicolumn{1}{|c|}{}                      &                              & \multicolumn{1}{l|}{IFFT}  & \multicolumn{1}{c|}{$\mathcal{O}(\log_2K)$}      &                                                                                         & \multicolumn{1}{l|}{SFFT}      & \multicolumn{1}{c|}{$\mathcal{O}(\log_2KM)$}     &                                                                                     \\ \cline{3-4} \cline{6-7}
\multicolumn{1}{|c|}{}                      &                              & \multicolumn{1}{l|}{MRT}   & \multicolumn{1}{c|}{$\mathcal{O}(M_t)$}                     &                                                                                         & \multicolumn{1}{l|}{Detection} & \multicolumn{1}{c|}{$\mathcal{O}(1)$}                       &                                                                                     \\ \cline{2-8}
\multicolumn{1}{|c|}{}                      & \multirow{2}{*}{Zak-based}   & \multicolumn{1}{l|}{IDZT}  & \multicolumn{1}{c|}{$\mathcal{O}(M_t\log_2M)$}     & \multirow{2}{*}{$\mathcal{O}(M_t\log_2M)$}                       & \multicolumn{1}{l|}{DZT}       & \multicolumn{1}{c|}{$\mathcal{O}(\log_2M)$}     & \multirow{2}{*}{$\mathcal{O}(\log_2M)$}                           \\ \cline{3-4} \cline{6-7}
\multicolumn{1}{|c|}{}                      &                              & \multicolumn{1}{l|}{MRT}   & \multicolumn{1}{c|}{$\mathcal{O}(M_t)$}                      &                                                                                         & \multicolumn{1}{l|}{Detection} & \multicolumn{1}{c|}{$\mathcal{O}(1)$}                       &                                                                                     \\ \hline
\multicolumn{1}{|c|}{\multirow{3}{*}{DDAM}} &  MRT                          & \multicolumn{1}{l|}{path-based MRT}   & \multicolumn{1}{c|}{$\mathcal{O}(M_tL)$}                               & $\mathcal{O}(M_tL)$                                                                  & \multicolumn{1}{l|}{Detection} & \multicolumn{1}{c|}{$\mathcal{O}(1)$}                       & $\mathcal{O}(1)$                                                                  \\ \cline{2-8}
\multicolumn{1}{|c|}{}                      & ZF                           & \multicolumn{1}{l|}{path-based ZF}    & \multicolumn{1}{c|}{$\mathcal{O}(M_tL^2/N_s+M_tL)$}                      & $\mathcal{O}(M_tL^2/N_s+M_tL)$                                                         & \multicolumn{1}{l|}{Detection} & \multicolumn{1}{c|}{$\mathcal{O}(1)$}                       & $\mathcal{O}(1)$                                                                  \\ \cline{2-8}
\multicolumn{1}{|c|}{}                      &  MMSE                         & \multicolumn{1}{l|}{path-based MMSE}  & \multicolumn{1}{c|}{$\mathcal{O}(M_t^3L^3/N_s+M_tL)$}                      & $\mathcal{O}(M_t^3L^3/N_s+M_tL)$                                                         & \multicolumn{1}{l|}{Detection} & \multicolumn{1}{c|}{$\mathcal{O}(1)$}                       & $\mathcal{O}(1)$                                                                  \\ \hline
\end{tabular}
}
\footnotemark[1]$M_t$: number of transmit antennas, $K$: number of OFDM subcarriers, $M$: number of OFDM symbols per OTFS frame, $L$: number of multi-paths, $N_s$: number of total transmit information-bearing symbols per channel coherence block.\\
\end{table*}

\subsection{Extensions of DDAM: DDAM + ``X''}\label{DDAM_X}
DDAM and other waveforms are not {\it exclusive}.
As illustrated in Fig.~\ref{DDAM}, we introduce a new framework for future waveform design, i.e., {\it combining} DDAM with other waveforms for further performance enhancement, which is referred to as {\it DDAM + ``X''}, where {\it ``X''} could be existing single- or multi-carrier waveforms.
In the following, we discuss two such examples, namely, {\it DDAM-OFDM} and {\it DDAM-OTFS}.

{\it DDAM-OFDM}:
The performance of conventional OFDM is is determined by the propagation channel.
In contrast, by combing DDAM with OFDM, DDAM-OFDM transmits over the resulted equivalent channel after applying DDAM, which has the reduced channel delay and Doppler spreads, thus enjoying the quasi-static flat fading channel.
With the reduced channel delay spread, the PAPR can be reduced by using fewer subcarriers, while the higher SE can be achieved as the CP overhead can be saved.
For DDAM-OFDM, as illustrated in Fig.~\ref{DDAM}, a frequency-domain beamforming is first performed before the DDAM transmitter at the subcarrier-level.
Following that, the path-based beamforming is performed in the time-domain for DDAM.

{\it DDAM-OTFS}:
Traditional OTFS faces limitations in the frame format. When the channel delay and Doppler spreads are large, information symbols may experience aliasing in the DD domain.
Moreover, if the system has a stringent PAPR or SE requirement, it becomes challenging to find suitable frame format parameters for OTFS.
In contrast, combining DDAM with OTFS can alleviate the constraints on OTFS frame format selection by reducing the channel delay and Doppler spread.
Furthermore, decreasing channel delay and Doppler spread significantly reduces the complexity of OTFS equalization at the receiver.
For DDAM-OTFS, information symbols are still multiplexed in the DD domain and transformed from the DD domain to the time domain via OTFS modules.

\subsection{Main Design Considerations}
This subsection presents the main design considerations specifically for {\it DDAM + ``X''}.
The following three aspects are discussed, i.e., channel state information (CSI) acquisition, fractional delay-Doppler compensation, and joint multi-domain beamforming.

\subsubsection{CSI Acquisition for DDAM}
The prerequisite of DDAM is to accurately acquire the CSI.
Note that different from the conventional TF domain waveform design that requires to estimate the CSI in TF domain superimposed by all the multi-paths, the CSI acquisition for DDAM needs to extract the features of individual multi-paths in terms of propagation delay, Doppler frequency shift, and AoD/AoA, etc., thus referred to as {\it path state information} (PSI) \cite{xiao2024Exploiting}.
Thus, most existing channel estimation methods cannot be directly applied to estimate the PSI.
However, the channel estimation for DDAM aligns with the propagation environment scatterer sensing.
Thus, it is an appealing option to seamlessly integrate the channel estimation for DDAM with the environment sensing.
Moreover, as the individual channel path varies much slower than the composite channel, to estimate the PSI, in addition to the channel coherence time, a new timescale referred to as the {\it path invariant time} can be taken into account \cite{xiao2024Exploiting}, during which the PSI can be assumed to remain unchanged.

\subsubsection{Fractional Delay-Doppler Compensation}
In practice, due to the finite delay and Doppler resolutions, there exist fractional delays and Dopplers, where the power of individual channel path may leakage into the nearby DD taps.
There are two types of delay-Doppler compensation methods, namely, {\it tap-based} \cite{lu2023delay} and {\it path-based} \cite{xiao2024Exploiting}.
For the tap-based method, the delay-Doppler compensation is performed based on all dominant taps in the DD domain, where each tap contains multiple physical paths.
By contrast, for the path-based method, only one DD tap associated with the highest power of each path is selected to be aligned.
Compared to the tap-based method, the path-based method enjoys the low complexity for beamforming, while the tap-based method may achieve a better performance than the path-based alternative.
Nevertheless, neither of them can completely eliminate the ISI due to the fractional delay and Doppler issue.
One effective approach is to design the appropriate delay and Doppler windows, which aims to reduce the channel delay and Doppler spreads within the desired value instead of completely eliminating them.

\subsubsection{Joint Multi-domain Beamforming}
The combination of DDAM with other waveforms, such as OFDM and OTFS, entails the joint multi-domain beamforming design.
For instance, for DDAM-OFDM, although the path-based ZF beamforming can significantly reduce the channel delay and Doppler spreads, there may still exist some residual ISI due to the fractional delays and Dopplers.
Thus, the frequency-domain beamforming can be performed at each subcarrier of OFDM to mitigate the residual ISI.
However, although the path-based ZF beamforming can effectively eliminate the ISI, it restricts the spatial DoFs for the frequency-domain beamforming.
Thus, it leads to new optimization problems on the joint design of multi-domain beamforming for {\it DDAM + ``X''}.

\section{Future Research Directions}\label{directions}

\subsection{Waveform Design and Performance Analysis}
The waveform design of DDAM with implementation impairments, such as the CFO, synchronization error, and phase noise, is still an open problem.
Specifically, DDAM may suffer from channel estimation error and fractional delay.
Thus, how to design waveforms to mitigate such issues is still a critical issue.
Moreover, to fully understand its potential for future wireless networks, it is necessary to conduct a comprehensive performance analysis and comparison of DDAM with other waveforms, such as OFDM, OTFS and ODDM.
In addition, for practical implementations, the research on DDAM with hybrid beamforming is important to pursue further.

\subsection{Combining DDAM with Other Waveforms}
Our discussions in Section~\ref{DDAM_X} briefly introduce the potential advantages of DDAM-OFDM and DDAM-OTFS compared to the conventional OFDM and OTFS.
Based on them, a very interesting topic arises on how to combine DDAM with other waveforms.
As the DDAM technique can significantly reduce the channel delay and Doppler spreads, it brings additional design DoFs for the TF domain or DD domain waveform design.
However, the pulse shaping and time-frequency resources allocation with DDAM are still open problems.
Moreover, how to jointly optimize the waveform and beamforming schemes in delay, Doppler, and spatial domains is still a challenging task in general.

\subsection{DDAM Multiple Access}
The efficient multiple access scheme is critical for 6G.
The DDAM technique provides new opportunities to suppress the multi-user inference (MUI).
Specifically, the propagation delays and Doppler frequencies associated with each user can be aligned into a desired region, while the resulting MUI can be mitigated through path-based beamforming in the spatial domain.
By leveraging the DDAM technique, how to flexibly allocate the time-frequency-spatial domain radio resources for multi-users is a promising direction for future research.

\subsection{DDAM for ISAC}
DDAM technique can be exploited for ISAC in two common manners.
On one hand, as discussed in Section~\ref{pDDAM}, DDAM enjoys the advantages of low PAPR and robustness to Doppler frequency shift, rendering it appealing for wireless sensing.
Compared to OFDM, DDAM can achieve wider sensing coverage and higher accuracy due to its lower PAPR and more tolerance to Doppler frequency shift \cite{xiao2022waveform}.
In this regard, how to jointly design the path-based beamforming of DDAM for ISAC is critical.
On the other hand, as the goal of environment scatterer sensing coincides with that of channel estimation for DDAM, it is a viable approach to integrate both of them for ISAC.
Thus, a novel signal processing framework can be developed, where the sensed states of scatterers can be directly used for DDAM-based communication \cite{xiao2024Exploiting}, while the states of scatterers can be efficiently tracked during the DDAM-based communication and further used for updating the CSI.

\section{Conclusion}\label{clu}
In this article, starting from the fundamental constraints on OFDM parameter selection and 6G development trends, we introduced a new framework for future waveform design based on the DDAM technique, whereby the channel delay and Doppler spreads can be significantly reduced.
Two important examples of DDAM extensions, i.e., DDAM-OFDM and DDAM-OTFS, were also discussed.
Furthermore, the key design considerations for DDAM were elaborated.
Lastly, we pointed out some future research directions for DDAM and its extensions.
It is hoped that the new challenges and opportunities presented in this article will help pave the way for researchers to design innovative and more efficient waveforms for future wireless networks.

\bibliographystyle{IEEEtran}
\bibliography{6GWaveform}

\end{document}